

\def\l{\lambda}\def\m{\mu}\def\n{\nu}\def\r{\rho}\def\t{\tau}
\def\ha{{1\over 2}}
\def \na{\nabla}
\def\xp{x^+}\def\xm{x^-}
\def\gmn{g_{\m\n}}\def\ghmn{\hat g_{\m\n}}

\def\dx{\int d^2x\ \sqrt{-g}\ }
\def\dhx{\int d^2x\ \sqrt{-\hat g}\ }
\def \ma{{2 M\over \l}}
\def\dd{\partial_+}
\def\dm{\partial_-}
\def\dpm{\partial_\pm}
\def\af{asymptotically flat }
\def\st{spacetime }
\def\bh{black hole }
\def\tran{transformations }
\def\bhs{black holes }

\def\section#1{\bigskip\noindent{\bf#1}\smallskip}

\magnification=1200

\font\titolino=cmbx10
\font\tsnorm=cmr10
\font\tscors=cmti10

\font\tscorsp=cmti9
\magnification=1200

\def\PRD{{\tscors Phys. Rev. D }}

\def\NPB{{\tscors Nucl. Phys. B }}

\def\PLB{{\tscors Phys. Lett. B }}

\def\CQG{{\tscors Class. Quantum Grav. }}

\def\note{\advance\notenumber by 1 \footnote{$^{\the\notenumber}$}}
\def\ref#1{\medskip\everypar={\hangindent 2\parindent}#1}
\def\beginref{\begingroup
\bigskip
\leftline{\titolino References.}
\nobreak\noindent}
\def\endref{\par\endgroup}
\def\beginsection #1. #2.
{\bigskip
\leftline{\titolino #1. #2.}
\nobreak\noindent}
\def\beginack
{\bigskip
\leftline{\titolino Acknowledgments.}
\nobreak\noindent}
\def\begincaptions
{\bigskip
\leftline{\titolino Figure Captions}
\nobreak\noindent}
\nopagenumbers
\null
\vskip 5truemm
\rightline {INFNCA-TH9521}

\rightline { }
\rightline { }
\rightline{October 1995}
\vskip 15truemm
\centerline{\titolino TRACE ANOMALY AND HAWKING EFFECT IN GENERIC}
\bigskip
\centerline{\titolino 2D DILATON GRAVITY THEORIES}
\vskip 15truemm
\centerline{\tsnorm Mariano Cadoni}
\bigskip
\centerline{\tscorsp Dipartimento di Scienze Fisiche,}
\smallskip
\centerline{\tscorsp Universit\`a  di Cagliari,}
\smallskip
\centerline{\tscorsp Via Ospedale 72, I-09100 Cagliari, Italy.}
\smallskip
\centerline {\tscorsp and}
\smallskip
\centerline{\tscorsp INFN, Sezione di Cagliari.}
\bigskip
\vskip 15truemm
\centerline{\tsnorm ABSTRACT}
\begingroup\tsnorm\noindent
\baselineskip=2\normalbaselineskip
Black hole  solutions in the context of a generic matter-coupled
two-dimensional dilaton
gravity theory are discussed both at the classical and semiclassical level.
Starting from general assumptions, a criterion for the existence of \bhs
is given. The relationship between conformal anomaly and Hawking
radiation is extended to a broad class of two-dimensional dilaton gravity
models.
A general and simple formula relating the magnitude of the Hawking effect
to the dilaton potential evaluated on the horizon is derived.


\vfill
\leftline{\tsnorm PACS:04.70.Dy,11.25.-W, 97.60.Lf\hfill}
\smallskip
\hrule
\noindent
\leftline{E-Mail: CADONI@CA.INFN.IT\hfill}
\bigskip
\endgroup
\vfill
\eject
\footline{\hfill\folio\hfill}
\pageno=1
\beginsection 1. Introduction.
\smallskip
Two-dimensional (2D) dilaton gravity [1-10] is a subject that has been
intensively
investigated in the last years not only because of its intrinsic
mathematical interest but also because it provides a simple toy model
for studying the Hawking radiation of black holes.
Moreover the relation with string theory in noncritical dimensions and
the fact that gravity in two dimensions  is renormalizable could make
these models very useful in a full quantum description of physical black
holes.
Several models of 2D dilaton gravity have been analyzed in the
literature [5-10]. Most of them are more or less motivated either by the
relation with string theory or by the connection with 4D \bh physics.
We have for example the Callan-Giddings-Harvey-Strominger (CGHS) model
[5],
which has been used to describe backreaction effects in the \bh
evaporation process and the Jackiw-Teitelboim theory (JT) [6,7] that was
historically the first 2D dilaton gravity theory. Other models of current
interest are Spherically symmetric gravity (SSG) that is obtained by
retaining only the radial modes of 4D Einstein gravity  and string
inspired models that admit \bh solutions in 2D anti-de Sitter spacetime
[8,9].

Even though the spectrum of models for 2D dilaton gravity is large and
composite, a unified and complete description of the theory
still exists [1-4].
It turns out that dilaton reparametrizations and Weyl rescalings of the
metric relate different models in such way that the most general action
for 2D dilaton gravity depends only on a function of the dilaton field
(the potential). The resulting theory in its general form is simple and,
in absence of matter, is exactly solvable both at the classical and
quantum mechanical level. Though the classical structure of the theory,
including the solutions and the classical observables for black holes, is
well-understood, there are still two main unsolved problems.
First, in the generic model one can find, using arguments related to the
local structure of the spacetime, strong evidence for the existence of
\bhs [4].
A rigorous prove of the existence of \bhs implies a detailed knowledge of
the global structure of the spacetime. For the particular models mentioned
above one can explicitly show that \bhs do  exist but analogous
statements for the generic model are still lacking. In principle, being
the potential the only free input in the action, one should be able to
impose conditions on the functional form of the potential,
which would
assure that \bhs really exist.
Second, assuming that \bhs exist, one should be able to describe them
semiclassically. In particular one would like to use here the
well-known relationship between trace anomaly and Hawking effect [11].
Again, this has been done for some special cases [5,7,8,12] but a treatment
for the general model is still lacking.

In this paper we will focus on these two problems and we will find that
they are strongly related. We will set general conditions on the
functional form of the potential,
which will be enough to assure that \bhs
exist. On the other hand we will find, considering a matter-coupled dilaton
gravity theory, that the same conditions are crucial for having
a consistent semiclassical description of \bhs.
We will show that the relation between conformal
anomaly and Hawking radiation can be extended to  generic 2D dilaton
gravity introducing local, dilaton dependent, counterterm in the
semiclassical action. We will also derive a simple expression relating the
magnitude of the Hawking effect to the potential evaluated on the event
horizon.

The paper is organized in the following way.
In sect. 2 we briefly review the features of generic 2D dilaton gravity
that are relevant for our investigation. In sect. 3 we analyze the
global structure of the solution and we show how the functional form of
the potential can be constrained so that \bhs exist.
In sect. 4 we study the theory coupled to scalar matter fields.
In sec. 5 we discuss the relationship between conformal anomaly and
Hawking radiation in the context of  generic 2D dilaton gravity.
In sect. 6 we discuss some relevant special cases. Finally, in sect. 7 we
draw our conclusions.

\beginsection 2. Classical 2D dilaton gravity.
\smallskip
Our starting point is the most general two-dimensional action depending
on the metric $\ghmn$ and the dilaton $\phi$,
which is invariant under
coordinate transformations and contains at most two derivatives of the
fields. This action takes the form [1,2]
$$S[\ghmn ,\phi] ={1\over 2\pi}\dhx\left(D(\phi)\hat
R+\ha(\hat\nabla\phi)^2+
\l^2 W(\phi) \right),\eqno(2.1)$$
where $\hat R$ is the Ricci scalar and $D$ and $W$ are arbitrary
functions of the dilaton $\phi$. The model represents a generalization of
well-known 2D gravity theories such as the
CGHS model [5] and the Jackiw-Teitelboim
model [6]. In its general form,
given by (2.1), it has been already analyzed in the literature [1-4]. In
the following we will briefly review some basic features of the
model that are relevant for our further investigation.
The model defined by (2.1) actually depends only on the dilaton potential
$W(\phi)$, since the Weyl rescaling
$$\gmn=\exp\left(\ha\int{d\phi\over dD/d\phi}\right)\ghmn\eqno(2.2)$$
and the reparametrization of $\phi$, $\Phi=D(\phi)$, bring the
action (2.1) into the form:
$$S[\gmn ,\Phi] ={1\over 2\pi}\dx\left(\Phi R+
\l^2 V(\Phi) \right),\eqno(2.3)$$
where $V(\Phi)$ is an arbitrary function of $\Phi$.
The field equations derived from the action (2.3) have the
simple form:
$$R=-\l^2 {dV\over d\Phi},\eqno(2.4)$$
$$\na_\m\na_\n\Phi-{\l^2\over 2}\gmn V=0.\eqno(2.5)$$
One can show that a generalized Birkhoff's theorem holds for the
theory: for each choice of the potential $V$ and modulo spacetime
diffeomorphisms the general static solutions of the theory form a
one-parameter family of solutions.
In the Schwarzschild gauge the solutions can be written as following:
$$ds^2= -\left(J(\l r) -\ma\right)dt^2 +
\left(J(\l r) -\ma\right)^{-1}dr^2\eqno(2.6)$$
$$\Phi=\l r,\eqno(2.7)$$
where $J(\Phi)=\int^\Phi d\tau V(\tau)$.
The parameter $M$ labeling the solutions is a constant of motion,
 which can be
interpreted as the mass of the solution and can be expressed in
the  coordinate invariant form:
$$M=-{\l \over 2}\bigl((\na \Phi)^2-J(\Phi)\bigr).\eqno (2.8)$$
The solutions admit a Killing vector $k_\m$, whose magnitude is
$$k_\m k^\m=\ma -J(\Phi).\eqno(2.9)$$
If the equation
$$J(\Phi)=\ma\eqno(2.10)$$
admits at least one solution $\Phi_0$, such
that in a neighborhood of $\Phi_0$ the function $J(\Phi)$ is monotonic,
the Killing vector (2.9) becomes spacelike at $\Phi_0$, signalizing the
presence of an event horizon. One is then led to interpret the solution as
a black hole. But the existence of event horizons cannot be inferred
studying only local properties of the solution.
In the next section we will
discuss the global structure of the spacetime and we will show that for
a broad class of models the interpretation of (2.6) as a \bh is
possible.

In the next sections we will need the solutions (2.6), (2.7) written in
the conformal
gauge, $ds^2=-\exp(2\r) d\xp d\xm$. Fixing the residual gauge freedom
relative to the conformal subgroup of diffeomorphisms, the solutions can
be written as
$$e^{2\r}=\biggl(J-\ma\biggr),\eqno(2.11)$$
$$\int^\Phi{d\t\over {J(\t)-\ma}}={\l\over 2}(\xp-\xm).\eqno(2.12)$$
Assuming that the solutions represent \bhs one can associate to them
thermodynamical parameters. For the temperature of a generic
\bh one has
$$T={\l\over 4\pi}V(\Phi_0).\eqno(2.13)$$
\beginsection 3. Global structure of the spacetime.
\smallskip
As stated in the previous section a rigorous prove of the existence
of \bh involves a detailed analysis of the global structure of the
spacetime. Being the potential $V$ the only free input in the action (2.3)
it is
evident that the information about this global structure is encoded
in the particular form of the function $V(\Phi)$.  In the following,
starting from general conditions that assure the existence of \bhs
solution, we will single out a broad class of models for which the
interpretation of (2.6) as a \bh can be well established.
These conditions will be translated in some constraints about the
functional form of the potential $V$. We will consider for simplicity
only \bhs with a single event horizon. Our discussion can be easily
generalized to the case of multiple horizons.

A crucial role in our analysis is played by the field $\Phi$. Due to
its scalar character $\Phi$ gives a coordinate-independent notion of
location
and can be therefore used to define the asymptotic region, the
singularities and the event horizon of our 2D spacetime.
We will consider only the \st region for which $0\le\Phi<\infty$. This
restriction is justified by the fact that the natural coupling constant
of the theory is $g=(\Phi)^{-1/2}$; the \st can  therefore be divided in a
strong coupling region ($\Phi=0$) and a weak coupling asymptotic region
($\Phi=\infty$). $\Phi$ is the 2D analogous of the radial coordinate $r$ in
4D spherically symmetric solutions (this analogy is particularly evident
for SSG, where the area of the transverse two-sphere is proportional to
$\Phi$).

The next step in our analysis is to write down a set of  conditions
that, if fulfilled, make the interpretation of (2.6) as a \bh
meaningful. Let us assume that
\smallskip
\noindent{\tscors a)} The equation (2.10)  has only one solution for
$\Phi=\Phi_0 >0$ with $V(\Phi_0)\neq 0$ and $V(\Phi)>0$ for
$\Phi>\Phi_0$.
\smallskip
\noindent{\tscors b)} For $M>0$  naked singularities are not  present;
the states with $M<0$  describe naked singularities.
\smallskip
\noindent{\tscors c)}  In the  $\Phi=\infty$  asymptotic region the
Killing
vector (2.9)
is timelike for every finite value of the mass $M$.
\smallskip
\noindent{\tscors d)} In the  $\Phi=\infty$ asymptotic region the curvature
 $R$ is finite.

\noindent Condition {\tscors a)} is necessary for the presence of an event
horizon.
It implies that the Killing vector (2.9) is timelike for $\Phi>\Phi_0$
and becomes spacelike for $\Phi<\Phi_0$.
Condition {\tscors b)} is necessary for the existence of \bhs for every
positive value of the mass and to assure that the vacuum $M=0$ has no
event horizons, it implies that  for every curvature
singularity $\Phi_1$  we have $\Phi_1<\Phi_0$ and that
the function $J(\Phi)$ has no zeroes.
Both conditions translate in very weak constraints on the functional form
of
$V$. Conditions {\tscors c)} and {\tscors d)} constrain strongly the
asymptotic
behavior of $V$. In fact {\tscors c)} implies that $J(\Phi)\to\infty$ as
$\Phi\to\infty$. This can be easily demonstrated, {\tscors ab assurdo},
assuming
that $J(\Phi=\infty)=l$ with $l$ finite. If this is the case for $M>{\l l/
2}$
the Killing vector (2.9) becomes spacelike. Furthermore condition {\tscors d)}
determines the degree of divergence of $J$. By looking at equation (2.4)
one
easily realizes that if the curvature must stay finite as $\Phi\to\infty$
the  function $J$ must diverge lesser or equal then $\Phi^2$.
In conclusion \bhs do exist if the function $V$ behaves asymptotically as
$$V\sim \Phi^a, \qquad -1< a\le 1.\eqno(3.1)$$
We have correspondingly two classes of black holes:
for $-1< a <1$, $R\to 0$ as $\Phi\to \infty$.
for  $ a =1$ $R\to const$ as $\Phi\to \infty$.
The asymptotic behavior $R\to 0$ is not enough to assure that the \st is
asymptotically flat.
Asymptotic flatness requires that asymptotically the metric can be put
in a Minkowski form. This issue will be settled at the end of this
section after the analysis of the global structure of the spacetime.

At this point the alert reader could object that the interpretation of
(2.6) as a \bh fails even though conditions {\tscors a)-d)} are fulfilled,
if
the \st has no curvature singularities and can be maximally extended to
describe a regular spacetime. This   happens, for example, for $V(\Phi)=1$
[12] and  for the JT theory ($V(\Phi)=2\Phi$) [7].
In these models the \bh spacetime can be maximally extended to become the
whole of Minkowski and anti-de Sitter spacetime respectively for the two
cases.
As pointed out in [7,12] if one cuts the \st at the line $\Phi=0$ (as we
do here) this extension is not possible and the \st does indeed represent
a black hole.  For the general model, in absence of curvature
singularities,
we will therefore consider the line $\Phi=0$ as the
boundary of the spacetime.

Two-dimensional dilaton gravity models in which the potential behaves
like (3.1) with arbitrary $a$, have been already discussed in ref. [13].
In that paper the author pointed out that the solutions with $a<1$ and
$a>1$
present essentially the same physical behavior if one interchanges
$\Phi=0$
with $\Phi=\infty$. Here, we will not consider this possibility because
we want to maintain the identification of the asymptotic region with the
weak coupling region $\Phi=\infty$.

The previous discussion enables us to single out
those models in (2.3) for
which  the solutions (2.6) can be consistently interpreted as black holes.
A detailed description of the causal structure of the \st depends of course
on the specific form of the potential $V$. The form of the Penrose
diagram will depend on the presence of single or multiple horizons and
on the nature of the singularities. Yet, the knowledge of the
asymptotic  behavior of the \st enables us to
infer some general
conclusions about the causal structure of the \st and to draw a
qualitative Penrose diagram.
For our class of models the metric behaves asymptotically as
$$ds^2= -(\l r)^{a+1} dt^2+(\l r)^{-(a+1)}dr^2, \quad -1<a\le 1,\quad 0\le
r<\infty ,\eqno(3.2)$$
where we made use of equation (2.7).
Performing the coordinate \tran $|a|\l y=(\l r)^{-a}, \quad\xp=t+y,
\quad \xm=t-y$, the
metric becomes
$$ ds^2= -\biggl({|a|\l\over 2}(\xp-\xm)\biggr)^{-(a+1)/a}d\xp d\xm.$$
{}From this expression of the metric one can read off the Penrose
diagram,
which turns out to depend on the value of $a$.
For $-1<a\le 0$ the metric can be put asymptotically in a Minkowski form.
The \st is asymptotically flat, the line $r=\infty$ ($\Phi=\infty$) is
lightlike
whereas $r=0$ ($\Phi=0$) is timelike. The Penrose diagram is shown in Fig.
1.
For $0<a< 1$, $R\to 0$ as $r\to \infty$ but the metric singularity at
$r=\infty$ ($\xp=\xm$) cannot be
removed by any coordinate  transformation.  We have the strange situation in
which though the  curvature is asymptotically zero the metric cannot be
put
asymptotically in a  Minkowski form.
The line $r=\infty$ is timelike whereas $r=0$ is lightlike.
The penrose diagram is depicted in fig. 2.
The case $a=1$ is analogous to the previous one. The only difference is
that now $R \to -\l^2$ as $r\to \infty$ so that the \st is asymptotically
anti-de Sitter.
Typical Penrose diagrams for \bhs with a single event horizon and with
the asymptotic behavior (3.1) are depicted
in Figs. 3 and 4.

We end this section with a brief discussion of the ground state of
the theory.
Generally we regard the solutions with $M=0$ as the ground state of the
theory. Condition {\tscors b)} assures that this state does not describe a
black hole,
but it does not guarantee that it describes a regular spacetime.
For example, for $V(\Phi)=1$ it has been shown that the vacuum is not a
regular \st but Minkowski space endowed with a null boundary [12]. At the
semiclassical level this fact poses non trivial questions about the
stability of the ground state [12]. Here we will follow the same approach
as in
[12], i.e. we will use a {\tscors cosmic censorship} conjecture to rule
out from
the physical spectrum the states of negative mass.

\beginsection 4. Coupling to (conformal) matter.
\smallskip
Two-dimensional dilaton gravity has no propagating degrees of freedom.
If one wants to describe a dynamical situation in which a \bh forms and
then (at the semiclassical level) eventually evaporates, one has to
couple the gravity-dilaton sector to matter fields.
We will consider here the simplest case of $N$, conformally coupled,
scalar matter fields $f$. The classical action is
$$S[\gmn ,\Phi,f] ={1\over 2\pi}\dx\left(\Phi R+
\l^2 V(\Phi)-\ha \sum^N_{i=1}(\na f_i)^2 \right).\eqno(4.1)$$
In the conformal gauge the equation of motion and the constraints
that follow from this action are
$$\eqalign{\dd\dm\r &=-{\l^2\over 8} e^{2\r}{dV\over d\Phi},\cr
\dd\dm\Phi &=-{\l^2\over 4} e^{2\r}V,\cr
\dd\dm f_i&=0,\cr
\dpm^2\Phi-&2\dpm\r\dpm\Phi=-T_{\pm\pm}^f,\cr}\eqno(4.2)$$
where $T_{\pm\pm}^f=\ha \sum^N_{i=1}(\dpm f_i)^2$ is the classical
energy-momentum tensor for the matter fields.
For generic $V$ and $T_{\pm\pm}$ this system of differential equation is
very hard to solve. Still a solution can be found,  maintaining a
generic $V$,   when we have only incoming matter in the form of
a shock-wave of magnitude $M$ at $\xp=\xp_0$, described by
$$T_{++}^f=M\delta(\xp-\xp_0),\quad T_{--}^f=0.$$
Thanks to Birkhoff's theorem we can find the solution simply by patching
together along the trajectory of the shock-wave a vacuum solution
( (2.11 ) and (2.12) with $M=0$) with a \bh solution. We have
$$\eqalign{e^{2\r}&=J,\cr
\int^\Phi{d\t\over J(\t)}&={\l\over 2}(\xp-\xm),\cr
\hbox{for} \qquad& \xp\le\xp _0.\cr}\eqno(4.3)$$
and
$$\eqalign{e^{2\r}&=\biggl(J-\ma\biggr)F'(\xm),\cr
\int^\Phi{d\t\over {J(\t)-\ma}}&={\l\over 2}\left(\xp-\xp _0
-F(\xm)\right),\cr
\hbox {for}\qquad &\xp\ge\xp _0,\cr}\eqno(4.4)$$
%
where
$$F'(\xm)={dF\over d\xm}=\biggl({J\over
{J-\ma}}\biggr)_{\xp=\xp_0}.\eqno(4.5)$$
Notice that the form of the function $F$ is such that both $\r$ and
$\Phi$ are continuous along the line $\xp=\xp_0$.
\beginsection 5. Trace anomaly and Hawking radiation.
\smallskip
So far our discussion has been purely classical. In a first approximation
one can describe quantum effects at the semiclassical level, by quantizing
the
matter fields in the fixed classical background of a \bh formed by
collapsing matter. In two dimensions this semiclassical description is
greatly simplified by the relation between conformal anomaly and Hawking
radiation discovered in ref. [11]. In the context of 2D dilaton gravity
this
relation has been already used to study the evaporation of the CGHS
\bhs [5] and of \bhs in anti-de Sitter \st [8]. The generalization to
 a generic theory of 2D dilaton gravity is still
not trivial. In fact  using the one-loop conformal anomaly
contribution stemming from the usual Liouville-Polyakov term one obtains
an expression for the quantum corrected energy-momentum tensor
$<T_{\m\n}^f>$ that is different from zero when evaluated on the ground
state of the theory.
Besides a simple calculation shows that, for the class of models
discussed in sect. 3, $<T_{\m\n}^f>_{gs}\sim \Phi^{2a}$  asymptotically.
For $a>0$ the semiclassical energy-momentum tensor diverges in the
$\Phi\to \infty$
asymptotic  region. At first glance this behavior
seems to make it impossible to use conformal anomaly arguments for the
study of the Hawking effect.
This conclusion is  strictly true only if one assumes that  the
contributions to the trace anomaly come entirely from the usual
nonlocal Polyakov-Liouville action. In general we have the freedom
to add local, covariant, dilaton-dependent counterterms to the
semiclassical action. The presence of dilaton-dependent contribution
to the  trace anomaly is natural if one treats
the metric and the dilaton on the same footing and is  crucial
if one wants to relate
the trace anomaly to the Hawking effect. The inclusion of such terms in
the
semiclassical action has been already discussed in the literature.
 In particular Strominger [14] has observed that the form of
the Polyakov anomaly action depends on the metric used to define
the path integral measure. The form of the counterterms can be
chosen in order to fulfill some physical conditions, for instance
that they respect all the symmetries of the classical theory [15],
that the theory becomes a conformal field theory [16]
or that the reparametrization ghosts decouple from the outgoing
energy flux [14]. Here we will follow an approach similar to the one
used in ref. [12] where the form of the dilaton-dependent counterterms
were determined by imposing physical conditions on the semiclassical
energy-momentum tensor evaluated on the classical vacuum of the theory.

Let us now consider the general form of the semiclassical action:
$$S=S_{cl}-{N\over 96\pi}\dx\left[ R{1\over \nabla^2} R
-4 H(\Phi) R +4G(\Phi) (\nabla\Phi)^2\right],\eqno(5.1)$$
where $S_{cl}$ is the classical action (4.1) and $H(\Phi),G(\Phi)$
are two arbitrary functions.
The first term above is the usual nonlocal
Liouville-Polyakov  term (we set to zero the contribution of the  ghosts)
whereas the other two represent the most general local, covariant,
with no second derivatives
counterterms one can add to the semiclassical action  (we do not consider
 modifications of the cosmological constant
action, i.e terms of the type $\lambda^2 N(\Phi)$, because they are
irrelevant for our considerations).

The Liouville-Polyakov term in the action (5.1) becomes local
in the conformal gauge, so that in this gauge
one can easily derive the quantum
contributions of the matter fields  to
the energy-momentum tensor, we have:
$$ <T_{+-}^f>=-{N\over 12}\biggl( \partial_+\partial _-\rho+
\partial_+\partial _- H\biggr),\eqno (5.2)$$
$$ <T_{\pm\pm}^f>=-{N\over 12}\biggl (\partial_\pm\rho\partial_\pm
\rho-
\partial_\pm^2 \rho +2 \partial_\pm\rho \partial_\pm H
-\partial_\pm^2 H- G \partial_\pm\Phi\partial_\pm
\Phi +t_\pm(x^\pm) \biggr).\eqno (5.3)$$
The functions $t_\pm(x^\pm)$ reflect the nonlocal nature of the
anomaly and are determined by boundary conditions.
Next, we have to find the form of the functions $H$ and $G$.
They can be fixed by the condition that the  energy-momentum tensor
vanishes
identically when evaluated on the classical ground state of the theory:
$$<T_{\m\n}^f>_{gs}=0.\eqno(5.4)$$
Assuming that the functions $t_{\pm}$ vanish for the ground state and
using eqs. (5.2), (5.3), (4.3), from (5.4) we get
$$\eqalign {H(\Phi)&=-{\ha}\ln J(\Phi)+ c\int^\Phi{d\t\over J(\t)},\cr
G(\Phi)&=-{1\over 4}{1\over J^2(\Phi)}\biggl({d J\over d\Phi}\biggr)^2+
 {c\over J^2(\Phi)}{d J\over d\Phi}.\cr}\eqno(5.5)$$
The constant $c$ appearing in the previous equations is arbitrary. It
turns out that the Hawking radiation effect does not depend on $c$ so
that we can take, without loss of generality, $c=0$.

The expressions (5.2) and (5.3) for the energy-momentum tensor are now
unambiguously
determined and we can turn to the calculation of the Hawking radiation
from
a \bh formed by collapse of a $f$ shock-wave as in (4.3), (4.4).
For $\xp\le \xp_0$ the solution is given by the vacuum (4.3) so that we
have
$<T_{\m\n}^f>=0$. For $\xp\ge \xp_0$ the solution is
given by the \bh solution (4.4) so that using (5.5) the expressions (5.2)
and
(5.3) become respectively:
$$<T_{++}^f>= -{N\l M\over 48}\biggl[{1\over J}\biggl(1-{M\over \l}{1\over
J}
\biggr)
\biggl({dJ\over d \Phi}\biggr)^2-\biggl(1-{2M\over \l}{1\over J}\biggr)
{d^2J\over d \Phi^2}\biggr],\eqno(5.6)$$
$$<T_{--}^f>= (F')^2<T_{++}^f>+{N\over 24}\{F,\xm\},\eqno(5.7)$$
$$<T_{+-}^f>=-{N\l M\over 24}\biggl[{1\over J}\biggl(1-{M\over \l}{1\over
J}
\biggr)
\biggl({dJ\over d \Phi}\biggr)^2-\ha \biggl(1-{2M\over \l}{1\over J}\biggr)
{d^2J\over d \Phi^2}- {1\over 2J}{d J\over d\Phi}\biggr],\eqno(5.8)$$
where $F$ is given by (4.5) and $\{F,\xm\}$ denotes the Schwarzian
derivative of the function $F(\xm)$.
Using the expression (3.1) for the asymptotic behavior of the function
$V(\Phi)$ one can now read off the values of the energy momentum tensor in
the
asymptotic region by taking the limit $\Phi \to \infty$. For \af spaces
($-1<a\le 0$) this limit can be taken in two different ways, either
$\xp\to
\infty$ or $\xm\to -\infty$. Because we are interested on the value taken
by
the energy-momentum tensor in the future null infinity region
we will let $\xp\to \infty$ as $\Phi\to \infty$ at fixed $\xm$.
For $0<a\le 1$ the line $\Phi=\infty$ is timelike and can be reached by
letting $\xp\to\xm$. In both cases the result of the limit will be a
function of the retarded coordinate $\xm$ and
will depend on the value of the parameter $a$ that characterizes
the asymptotic behavior of the spacetime. For $-1<a<1$  we
have:
$$ <T_{++}^f>\to 0,\qquad <T_{+-}^f>\to 0,\eqno(5.9)$$
$$<T_{--}^f>\to {N\over 24}\{F,\xm\},\eqno(5.10)$$
For $a=1$ the spacetime is asymptotically anti-de Sitter and
$<T_{++}^f>\to A$, with $A$ constant. The constant term in the
asymptotic expression for  $<T_{++}^f>$ and $ <T_{--}^f>$ can be
eliminated with an appropriate choice of the functions $t_\pm$ appearing
in (5.3). Setting $t_+(\xp)=(12/N)A\theta(\xp-\xp_0)$ and
$t_-(\xm)=(12/N)A\theta(\xm-\xp_0)$ we obtain also for $a=1$ the same
results (5.9), (5.10). On the other hand we still have $<T_{+-}^f>\to
const.$ This
constant term can be interpreted as the quantum correction to the vacuum
energy of the anti-de Sitter background.
It is important to notice that the asymptotic behavior (3.1) of the
potential $V$ is crucial for having a well-behaved expression for
$<T_{\m\n}^f>$. For example if $V$ behaves asymptotically as in (3.1) but
with
$a>1$, $<T_{\m\n}^f>$ will diverge as $\Phi\to \infty$.
Thus equation (3.1) not only takes care that \bhs do exist but also
assures
that a semiclassical description of them is possible.

The limiting value  $<T_{--}^f>_{as}$ in (5.10) is the flux of
$f$-particle energy
across future  infinity. However insertion of $F$ given by (4.5) in (5.10)
shows that   $<T_{--}^f>_{as}$ diverges as the horizon is approached.
This is due to the bad behavior of our coordinate system on the horizon.
The divergence can be easily eliminated by defining the new light-cone
coordinate $\hat \xm=F(\xm)$, with $F$ given by eq. (4.5).
Because in (5.3) we have set $t_-=0$, $<T_{--}^f>$ transforms anomalously
under conformal coordinate transformations.
Using the anomalous transformation law of
$<T_{--}^f>$  (see for example [8]), one finds
$$ <\hat T_{--}^f>_{as}={N\over 24}{1\over (F')^2}\{F,\xm\}.\eqno(5.11)$$
This expression is well-behaved on the horizon. Inserting eq. (4.5) into
(5.11)
we find that as the horizon $\Phi=\Phi_0$ is approached the Hawking flux
reaches the
constant (thermal) value:
$$ <\hat T_{--}^f>_{as}^h= {N\over 12}{\l^2\over 16}
\bigl[V(\Phi_0)\bigr]^2.\eqno(5.12)$$
This is the main result of this paper and is consistent with naive
thermodynamical arguments based on the formula (2.13) for the temperature
of
the black hole. In fact using (2.13) one can express the magnitude of the
Hawking
effect (5.12) as a function of the temperature, one has
 $$<\hat T_{--}^f>_{as}^h= {N \pi^2\over 12} T^2.$$

\beginsection 6. Special cases.
\smallskip
The general model (2.3) with the potential $V$ satisfying the conditions
discussed in sect. 3 contains, as particular cases, models that have
been already investigated in the literature both classically and
semiclassically. In this section we will show how previous results
on the Hawking effect can be obtained as particular cases of formula
(5.12).
Also we will work out an example of a model that admits \bh solutions
with multiple horizons.
\smallskip
\leftline{\tscors  String inspired dilaton gravity}
\smallskip
This is the most popular 2D dilaton gravity model. In its original
derivation [5], due
to CGHS, the action has the form (2.1).
The Weyl-rescaled model is of the form (2.3) with $V(\Phi)=1$.
The model admits \af \bh solutions [12]. Using formulae (2.13) and (5.12)
we find for the
temperature and magnitude of the Hawking effect
$$T={1\over 4\pi}\l, \qquad  <\hat T_{--}^f>_{as}^h={N\over 12}{\l^2\over
16}.\eqno(6.1)$$
This result coincides, after the redefinition $\l\to 2\l$ needed to match
the conventions of ref. [5,12], both with the CGHS results [5] and with
the
result of ref. [12] for the Weyl-rescaled model.
\smallskip
\leftline {\tscors Spherically symmetric gravity}
\smallskip
This model is obtained by retaining only the radial modes of
4D Einstein gravity.
It is characterized by $V(\Phi)=1/\sqrt{2\Phi}$ [4].
According to our classification of sect. 3 we have $a=-1/2$,
therefore the model admits \af \bh solutions with an event
horizon at $\Phi_0=2M^2/\l^2$. The corresponding Penrose diagram is
that shown in fig. 3. Using eqs. (2.13) and (5.12) we obtain
$$T={1\over 8\pi}{\l^2\over M}, \qquad  <\hat T_{--}^f>_{as}^h=
{N\over 12}{\l^4\over
64M^2}.\eqno(6.2)$$
The \bhs of this model have negative specific heat.
\smallskip
\leftline {\tscors The Jackiw-Teitelboim theory}
\smallskip
The JT theory is obtained from the action (2.3) by taking
$V(\Phi)=2\Phi$ (we use the conventions of ref [7]).
Being characterized by $a=1$ the model admits \bhs with
 anti-de Sitter behavior. More precisely, as shown in [7],
the \bh spacetime is obtained from a particular
parametrization of 2D anti-de Sitter spacetime endowed with a
boundary. The \bh horizon is at $\Phi_0= \sqrt{2M/\l}$ and
eqs. (2.13) and (5.12) give now
$$T={1\over 2\pi}\sqrt{2M\l}, \qquad  <\hat T_{--}^f>_{as}^h=
{N\over 24}M\l.\eqno(6.3)$$
The same result for the Hawking radiation rate has been obtained
in ref. [7]  performing the canonical quantization of the scalar
 fields $f$ in the
anti- de Sitter  background  geometry. Notice that the \bhs have positive
specific heat, indicating the emergence of a stable state as the
mass of the hole goes to zero.
\smallskip
\leftline {\tscors 2D \bhs in anti-de Sitter spacetime}
\smallskip
The models discussed in ref.[8,9], characterized by the action
$$S[\ghmn ,\phi] ={1\over 2\pi}\dhx e^{-2\phi}\left(\hat R+{8 k\over k-1}
(\hat\nabla\phi)^2+
\l^2\right),\eqno(6.4)$$
with $-1<k\le 0$, admit \bh solutions in anti-de Sitter spacetime.
The CGHS and the JT models appear as limiting cases of this general class
of dilaton gravity models for $k=-1,0$ respectively.
The action (6.4) can be mapped by a Weyl rescaling of the metric of the
form
(2.2) into action (2.3) with
$$V(\Phi)= \Phi^{(k+1)/(1-k)}.\eqno(6.5)$$
For $-1<k\le 0$ the parameter $a$ characterizing the asymptotic
behavior of the \bh solutions verifies $ 0<a\le 1$.
The \bh solutions of
these models evidenciate the peculiar asymptotic
behavior described in sect. 3. The Penrose diagram relative to them is
represented in fig. 4. The event horizon of the \bh is at
$\Phi=\Phi_0=[4M/((1-k)\l)]^{(1-k)/2}$. The temperature and the flux of
Hawking radiation are
$$T={1\over 2\pi}\biggl ({2M\over 1-k}\biggr)^{(k+1)/2}
\biggl({\l\over 2}\biggr)^{(1-k)/2}, \quad  <\hat T_{--}^f>_{as}^h=
{N\over 48}\biggl ({2M\over 1-k}\biggr)^{(k+1)}
\biggl({\l\over 2}\biggr)^{(1-k)}.\eqno(6.6)$$
The result (6.6) coincides with that found in ref. [8] for the model
(6.4), after
the redefinition $\l\to \quad\sqrt{2/(1-k)}\quad\l$, needed to match the
conventions of
[8].
\smallskip
\leftline{ \tscors 2D Black holes with multiple horizons}
\smallskip
We conclude this section with a model that admits \bh solution with
two event horizons. Let us consider the action (2.3) with
$$V(\Phi)= 1-(C/\Phi)^2, \eqno(6.7)$$
where $C$ is an arbitrary positive constant. According to our general
classification of sect. 3 we will have \af \bh solutions.
 The solutions (2.6), (2.7) become now
$$\eqalign {ds^2&= -\left(\l r+{C^2\over \l r} -\ma\right)dt^2 +
\left(\l r+{C^2\over \l r} -\ma\right)^{-1}dr^2,\cr
\Phi&=\l r.\cr}\eqno(6.8)$$
For $M> \l C$ the solution (6.8) describes \bhs with a singularity at
$r=0$ and two horizons at $r=r_\pm=\l^{-2}(M\pm\sqrt{M^2-(C\l)^2})$.
The \bh becomes extremal for $M= \l C$.
The potential $V$ evaluated on the outer horizon is
$$V(\Phi_0^{out})
=1-\biggl({C\l\over{M+\sqrt{M^2-(C\l)^2}}}\biggr)^2.\eqno(6.9)$$
Both the temperature and the Hawking radiation rate decrease with the
mass of the hole and become zero in the extremal case.

\beginsection 7. Conclusions.
\smallskip
We have been able to give a unified description both at the classical and
semiclassical level of the \bh solutions  of a general 2D dilaton gravity
theory. A criterion for the existence of \bhs has
been formulated and the relationship between conformal anomaly and
Hawking radiation has been extended to a broad class of 2D dilaton gravity
models. In particular we could write down a very simple and
general formula relating the magnitude of the Hawking effect to the dilaton
potential evaluated on the horizon.
The price that we had to pay for achieving this general description is a
strong constrain on the functional form of the potential, in particular
on its asymptotic behavior. The conditions discussed in sect. 3 rely
very heavily on the form of the action (2.3), they are sensitive to the
Weyl rescaling (2.2) that brings the action into the form (2.1).
Though the global structure of the solutions does not change under
this transformation, local quantities such as the Ricci curvature do change
so that some conditions of sect. 3 should be reformulated when one
considers the  Weyl-rescaled model. It may therefore be possible that a
more general
description exists that takes in full account the equivalence of models
under Weyl
rescalings of the metric. It may also be possible that in such general
framework the existence of \bh solutions
could be derived
imposing weaker restrictions to the  form of the potential.
At the semiclassical level a description that takes in full account the
equivalence of models under Weyl rescalings would be even more opportune.
In sect. 6 we have seen that the Hawking radiation
rate for string inspired gravity and  for the models (6.4) does not
change under a Weyl rescaling of the form (2.2).
It would be very interesting to see if this fact is a peculiarity of these
models or a general feature of  2D dilaton gravity.
On the other hand our discussion left  completely aside the backreaction
of the geometry on the radiation and quantum gravity effects. Both are
expected to be crucial in order to understand the end-point of the
evaporation process. The inclusion of the backreaction makes the theory
very hard to solve at least in its general form. Up to now only for a
modified version of the CGHS model (the Russo-Susskind-Thorlacius model
[15])
an exact solution could be found.

\smallskip

\beginack

I thank S. Mignemi  for useful comments. This work was partially supported
by MURST.
\smallskip

\beginref

\ref [1] T. Banks and M. O'Loughlin, \NPB {\bf 362}, 649 (1991).

\ref [2] R. B. Mann, \PRD {\bf 47}, 4438 (1993).

\ref [3] D. Louis-Martinez and G. Kunstatter, \PRD {\bf 49}, 5227 (1994).

\ref [4] J. Gegenberg,  G. Kunstatter and  D. Louis-Martinez,
{\tscors Talk given at the Conference on Heat Kernels and Quantum Gravity},
Winnipeg, 1994, gr-qc 9501017.

\ref [5] C.G. Callan, S.B. Giddings, J.A. Harvey and A. Strominger,
\PRD {\bf 45}, 1005 (1992).

\ref [6] C. Teitelboim, in {\tscors Quantum Theory of gravity },
S.M. Christensen,
ed. (Adam Hilger, Bristol, 1984); R. Jackiw, {\tscors ibidem}.

\ref [7] M. Cadoni and S. Mignemi, \PRD  {\bf 51}, 4319 (1995).

\ref [8] M. Cadoni and S. Mignemi, \NPB {\bf 427}, 669 (1994).

\ref [9] J.P.S. Lemos and P.M. S\'a, \PRD {\bf 49}, 2897 (1994).

\ref [10] T. Banks and M. O'Loughlin, \PRD {\bf 48}, 698 (1993);
D.A. Lowe, M. O'Loughlin, \PRD {\bf 48}, 3735 (1993);
S.P. Trivedi, \PRD {\bf 47}, 4233 (1993);
K.C.K. Chan and R. B. Mann \CQG {\bf 12}, 1609 (1995).

\ref [11] S. M. Christensen, S. A. Fulling, \PRD {\bf 15}, 2088
(1977).

\ref [12] M. Cadoni, S. Mignemi \PLB (in press).

\ref [13] S. Mignemi, {\tscors Ann. Phys.} (in press), hep-th 9411153.

\ref [14] A. Strominger, \PRD {\bf 46}, 4396 (1992).

\ref [15] J.C. Russo, L. Susskind, L. Thorlacius, \PRD {\bf 46},
3444 (1992).

\ref [16] A. Bilal, C. Callan, \NPB {\bf 394}, 73 (1993);
S.P. de Alwis, \PLB {\bf 289}, 278 (1992).

\endref
\vfill
\break
\begincaptions
\bigskip
\noindent
Figure 1.
\smallskip \noindent
Penrose diagram of the \st (3.2) with $ -1<a\le 0$.
\bigskip\noindent
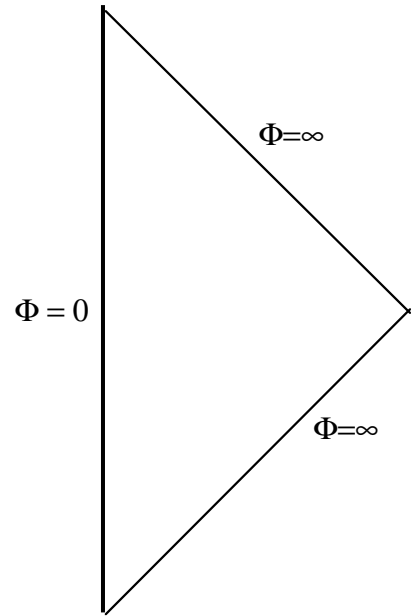
Figure 2.
\smallskip\noindent
Penrose diagram of the \st (3.2) with $ 0<a\le 1$.
\bigskip \noindent
Figure 3.
\smallskip \noindent
Typical Penrose diagram of a \bh with a single event horizon at $\Phi=\Phi_0$
 and with an
asymptotic behavior characterized by $ -1<a\le 0$.
\bigskip\noindent
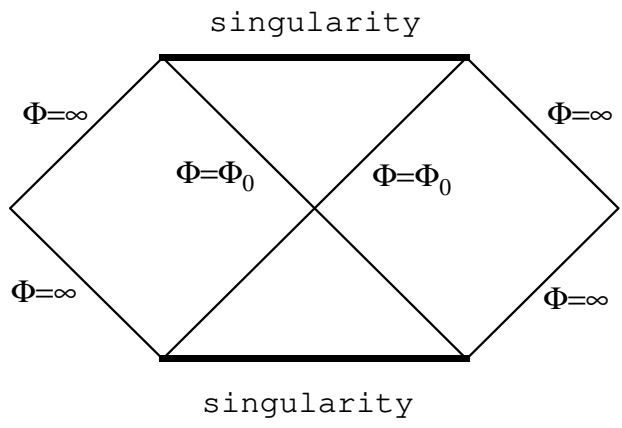
Figure 4.
\smallskip \noindent
Typical Penrose diagram of a \bh with a single event horizon at $\Phi=\Phi_0$
and with an
asymptotic behavior characterized by $ 0<a\le 1$.
\vfill
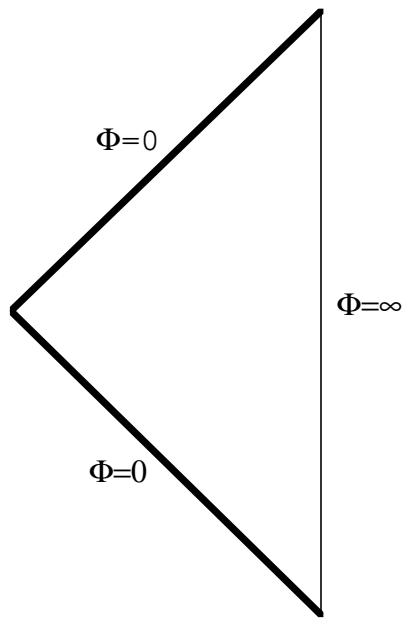
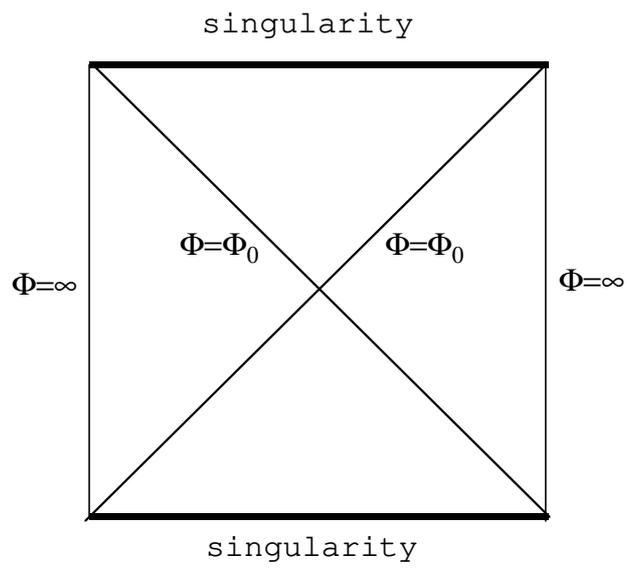
\end